\documentclass[fleqn, 10pt]{wlscirep}
\usepackage[utf8x]{inputenc}
\usepackage[T1]{fontenc}
\usepackage{lineno}

\usepackage{xcolor}
\usepackage[english]{babel}
\usepackage[colorlinks=true, allcolors=blue]{hyperref}
\usepackage{subcaption}
\usepackage{caption}
\usepackage{booktabs}
\usepackage{graphicx}
\usepackage{caption}
\usepackage{footmisc}
\usepackage{afterpage}
\usepackage{textalpha} 
\usepackage{float}
\usepackage{listings}
\usepackage{textgreek}

\newfloat{lstfloat}{htbp}{lop}
\floatname{lstfloat}{Listing}

\graphicspath{{figures/}{./}} 

\author[1,$\dag$]{Antonio Mirarchi}
\author[2,$\dag$,*]{Toni Giorgino}
\author[1, 3, 4, *]{Gianni De Fabritiis}

\affil[1]{Computational Science Laboratory, Universitat Pompeu Fabra, Barcelona Biomedical Research Park (PRBB), Carrer Dr.\ Aiguader 88, Barcelona, 08003, Spain.}
\affil[2]{Biophysics Institute, National Research Council (CNR-IBF),  Via Celoria 26, Milan, 20133, Italy}
\affil[3]{Institució Catalana de Recerca i Estudis Avançats (ICREA),  Passeig Lluis Companys 23, Barcelona, 08010, Spain.}
\affil[4]{Acellera Labs, Doctor Trueta 183, Barcelona, 08005, Spain.}

\affil[*]{corresponding author(s): Toni Giorgino (toni.giorgino@cnr.it) and Gianni De Fabritiis (g.defabritiis@gmail.com)}

\affil[$\dag$]{A. M. and T. G. contributed equally to this work}

\title{mdCATH: A Large-Scale MD Dataset for Data-Driven Computational Biophysics}

\begin{abstract}
Recent advancements in protein structure determination are revolutionizing our understanding of proteins. Still, a significant gap remains in the availability of comprehensive datasets that focus on the dynamics of proteins, which are crucial for understanding protein function, folding, and interactions. To address this critical gap, we introduce mdCATH, a dataset generated through an extensive set of all-atom molecular dynamics simulations of a diverse and representative collection of protein domains. This dataset comprises all-atom systems for 5,398 domains, modeled with a state-of-the-art classical force field, and simulated in five replicates each at five temperatures from 320 K to 450 K. The mdCATH dataset records coordinates and forces every 1 ns, for over 62 ms of accumulated simulation time, effectively capturing the dynamics of the various classes of domains and providing a unique resource for proteome-wide statistical analyses of protein unfolding thermodynamics and kinetics. We outline the dataset structure and showcase its potential through four easily reproducible case studies, highlighting its capabilities in advancing protein science.
\end{abstract}

\begin{document}

\flushbottom
\maketitle
\section*{Background and Summary}
Proteins, the building blocks of life, are central to nearly all biological processes, and understanding their structure and dynamics is crucial for advancements in fields ranging from biochemistry to pharmaceuticals. The convergence of advanced computational methods and biophysical techniques has led to unprecedented insights into molecular structures and functions of proteins. Molecular dynamics (MD), for example, is a compute-intensive technique that attempts to model the dynamics of biological macromolecules in realistic environments, often at all-atom resolution, based on empirical force-fields whose quality has been improving over decades \cite{piana_charmm22star, mackerell_all-atom_1998, piana_desamber_2020}. Machine learning, especially through the development of neural network potentials (NNPs), has the potential to further enhance computational protein research by enabling more accurate predictions and simulations of behaviors \cite{anand2022protein, mosalaganti2022ai, isert2023structure}. However, the lack of comprehensive datasets capturing the dynamic behaviors of proteins remains a significant challenge \cite{vander2021medusa}. Such datasets are vital for training machine learning models that can predict protein folding, functions, and interactions—often dynamic and transient processes, yet critical for understanding how macromolecules work, interact, and how they might be targeted. High-quality datasets are thus pivotal in advancing our comprehension of these complex phenomena. In recent years, efforts have been made to provide MD datasets, especially for key targets in drug discovery. Notable databases include GPCRmd \cite{rodriguez2020gpcrmd}, a platform dedicated to the study of G-protein-coupled receptors (GPCRs) dynamics, and SCOV2-MD\cite{scov2md_2022} as well as BioExcel-CV19 \cite{beltran2024new}, both showcasing the power of collaborative MD databases in the context of COVID-19 research. However, these initiatives are limited by their focus on specific proteome subsets, leaving a gap in comprehensive proteome-wide dynamic datasets. Previous projects such as MoDEL \cite{meyer_model_2010}, Dynameomics \cite{van_der_kamp_dynameomics_2010} and ATLAS \cite{vander2024atlas}, and the MDDB\cite{mddb} and MDRepo\cite{mdrepo} initiatives have been introduced to provide dynamics datasets encompassing a broader range of proteins, often in a single replica and at room temperature, but the computational cost of MD has generally limited databases in terms of coverage breadth and timescales. 

Here, we introduce mdCATH, a dataset focused on providing extensive all-atom MD-derived dynamics for most protein domains in the CATH classification system \cite{sillitoe_cath_2021}. mdCATH features simulations of 5,398 domains at five different temperatures, each in five replicas, therefore offering statistically relevant large-scale insights into protein structure dynamics under a multiplicity of conditions. This extensive and homogeneously-collected dataset of all-atom molecular dynamics simulations fills a critical void in the available molecular datasets by offering a rich, diverse, and physiologically relevant array of protein domain dynamics, enabling systematic, proteome-wide studies into protein thermodynamics, folding, and kinetics. It is possible to exploit mdCATH for learning data-driven (e.g. neural network-based) potentials \cite{doi:10.1021/acs.jctc.4c01239}, also thanks to the inclusion, unique to our knowledge, of instantaneous forces derived from a state-of-the-art all-atom force field. We hope that the mdCATH dataset will facilitate improvements in the design and refinement of biomolecular force fields.

\section*{Dataset requirements}
Our goal is to take a step forward in creating a proteome-wide molecular dynamics dataset for advancing drug discovery and enabling researchers to explore the dynamic behaviors of diverse protein targets. We built the mdCATH dataset to meet the following design features:

\begin{itemize}
\item \textit{Comprehensive coverage of structural features}. 
    mdCATH provides molecular dynamics information across 5,398 protein domains from the CATH classification system. This extensive coverage ensures a broad representation of the proteome, making the dataset valuable for a wide range of research applications in drug discovery.
\item \textit{MD-derived coordinates and forces}. 
    The dataset includes both coordinates and forces from simulated trajectories. The presence of forces is a unique feature in this dataset, which enables training force-based machine learning potentials.
\item \textit{Wide conformational space sampling}.
    mdCATH features multiple replicas at different temperatures, capturing a variety of conformations, including higher energy states encountered in molecular dynamics simulations. This ensures that the potential functions trained on this dataset produce accurate results across all relevant conformations.
\item  \textit{High quality data}.
    To ensure the highest accuracy, mdCATH utilizes state-of-the-art force fields, code, and computational resources. The accuracy of the dataset directly impacts the performance of models trained on it, making the use of the most accurate level of theory practical a priority.
\item \textit{Derived metadata}.
    The dataset includes pre-computed information such as root-mean-square deviation (RMSD), root-mean-square fluctuation (RMSF), secondary structure composition, and so on.
\item  \textit{Reproducibility}. 
    Reproducibility is ensured by including the PDB and PSF files in the dataset. Additionally, the data is stored in the efficient HDF5 binary data format, facilitating easy access and manipulation of the dataset for further research and model training.
\end{itemize}

\section*{Methods}

We built the dataset on the basis of the domain definitions provided by the CATH database \cite{cath_1, cath_2, cath_3}. CATH, a publicly available resource maintained by the Orengo group, provides a set of domains clustered by general architecture according to the class, architecture, topology, and homologous superfamily hierarchy \cite{sillitoe_cath_2021}.  We started from 14,433 non-homologous domains at the S20 (20\%) homology level in CATH release 4.2.0. We then restricted the selection to the subset of 13,470 domains between 50 and 500 amino acids, to focus on globular structures. Next, we excluded all the structures whose backbone was non-contiguous, e.g.\ due to unresolved regions in the original experimental structures; we also excluded sequences containing non-standard amino acids (also absent from CATH model files). The inclusion criteria left 5,883 residues for further processing. 

\begin{figure}[tb]
    \centering
    \includegraphics[width=.6\linewidth]{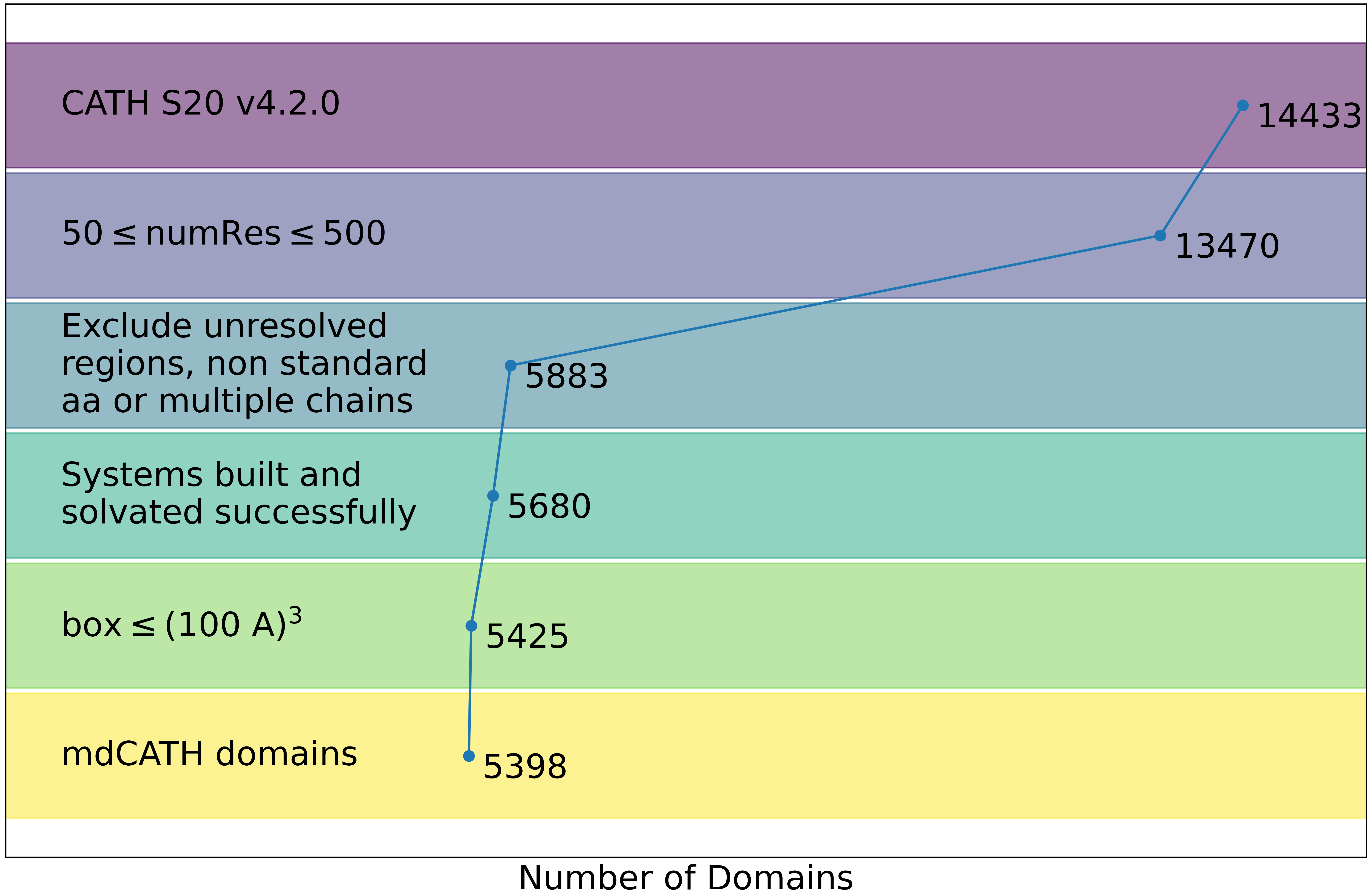}
    \caption{Exclusion criteria and the resulting number of domains at each step, starting from the 14,433 domains in the S20 homology set of CATH release 4.2.0, and ending with 5,398 domains included in the mdCATH dataset presented in this work.}
    \label{fig:exclusions}
\end{figure}

All the domain structures have been prepared with a standard protonation protocol at pH 7 including charge state assignments, proton placement and H-bond network optimization \cite{martinez-rosell_playmolecule_2017}.  Peptide chains were capped with acetylated and N-methylated termini. The systems were solvated in cubic boxes of TIP3P water with at least 9 \AA\ of padding on each side, neutralized, and ionized with Na$^+$ and Cl$^-$ ions at 0.150 M concentration. Systems whose resulting solvation cubic box was larger than (100 \AA)$^3$ were discarded. The final dataset includes 5,398 accepted domains, as illustrated in Figure~\ref{fig:exclusions}. HTMD version 1.16  was used for all the building steps \cite{doerr_high-throughput_2017, doerr_htmd_2016}.

All systems were parameterized with the CHARMM22* forcefield \cite{piana_charmm22star}. Long-range electrostatic forces were treated with the particle-mesh Ewald (PME) summation \cite{dardenParticleMeshEwald1993}, with an integration timestep of 4 fs enabled by the hydrogen mass repartitioning scheme of 4 amu per H atom \cite{feenstra_improving_1999}. The simulations were performed with ACEMD \cite{harvey2009acemd} on GPUGRID.net  distributed network \cite{buch_high-throughput_2010}.

Each system thus obtained was subjected to a pre-equilibration phase for 20 ns with a time-step of 4 fs in the NPT ensemble at 1 atm and 300~K utilizing the Montecarlo barostat. Harmonic restraints were applied to the protein's carbon α atoms  (1.0 kcal/mol/\AA) and heavy atoms (0.1 kcal/mol/\AA) to maintain them close to their initial positions during the first half (10 ns) of equilibration. The second half of equilibration (10 ns to 20 ns) was performed without restraints. No restraints were used during the subsequent production phase.

The final configuration of each system was used as a starting point for 25 production simulations, spawning runs at five temperatures in geometric progression (320 K, 348 K, 379 K, 413 K, 450 K), each in five replicas. The production simulations were performed in the NVT ensemble using Langevin thermostat for integration and a 0.1~ps$^{-1}$ relaxation time.
The use of the constant-volume ensemble sidesteps issues with the poor reproduction of the water phases and pressure by TIP3P \cite{quoikaLiquidVaporCoexistence2024,vegaWhatIceCan2008}.
Bonds involving hydrogen atoms were constrained at the equilibrium length with the M-shake algorithm \cite{krautlerFastSHAKEAlgorithm2001} with a tolerance of $10^{-5}$. Atom positions and forces acting on each atom were recorded every 1 ns and made available as part of the dataset as described below. A sampling rate of 1 ns bounds the tractable kinetics, enabling the resolution of the dynamics of relatively slow degrees of freedom such as conformational changes, but not faster motions (e.g. solvent-exposed side-chain rotations).
For both NPT and NVT simulations, a 9 \AA\ cutoff was applied for PME, while van der Waals interactions used a cutoff of 9 \AA\ and a switching distance of 7.5 \AA.
Analysis of the trajectories was conducted using the HTMD library\cite{doerr_htmd_2016}, in order to include potentially useful pre-computed metadata. Secondary structure assignments have been computed for each frame and residue using the implementation of the DSSP algorithm in \textit{moleculekit} version 1.8.32, encoded following the customary 8-class codes \cite{kabsch_dictionary_1983}. 

\section*{Data Records}

The mdCATH dataset makes the trajectories available under a  CC BY 4.0 license. It is available at HuggingFace \cite{mdcath_hf}.  It is possible to (1) download individual domain files from HuggingFace via a browser; (2) retrieve them via the HuggingFace dataset API (Listing \ref{lst:hf_api_mdcath}); (3) visualize them interactively (without downloading) on the PlayMolecule website (see the ``Code Availability'' section); (4) download them from PlayMolecule in  XTC format.


\subsection*{Organization}
The dataset is provided as a set of files in the Hierarchical Data Format, version 5 (HDF5). HDF5 allows the efficient storage and random access of heterogeneous data fields and arrays organized in a filesystem-like hierarchy. For the sake of simplicity, all of the data related to a given domain were collected into an individual HDF5 file. The dataset provided is structured into fields that describe snapshots of molecular simulation trajectories and derived quantities as shown in Table \ref{tab:h5}. The root group of each file in the dataset is the domain ID, which aggregates fields such as {\tt chain}, {\tt element},  {\tt resid}, {\tt resname}, and {\tt z},  each a vector of length $N$, representing the number of protein atoms. The \texttt{pdb} and \texttt{psf} strings hold, respectively, the verbatim PDB file used for the simulation (with solvent) and its topology in CHARMM/XPLOR protein structure file (PSF) format; \texttt{pdbProteinAtoms} holds a PDB of the $N$ solute atoms used for analysis. Data on the dynamics are organized hierarchically: five groups at the top-most level named according to the temperature; each temperature group includes five groups for each of the replicas; finally, each replica holds fields for atomic coordinates, forces, simulation box, as well as pre-computed derived quantities such as secondary structure assignments, instantaneous gyration radius, root-mean-square deviation, and fluctuations. Coordinates and forces are stored as three-dimensional arrays, their axes running along frames, atoms, and spatial dimensions. DSSP secondary structure assignments are provided per residue and frame following the standard 8-letter codes.

\begin{table*}[p]
\centering \small
\begin{tabular}{lcccl} 
\toprule
Field & Size & Type & Unit & Description \\
\midrule
\textit{Domain ID}\texttt{/}     & & & & \\
\quad \texttt{chain}   & $N$ & string &  &  Chain ID  \\
\quad \texttt{element} & $N$ & string & & Chemical element \\
\quad \texttt{pdb}     & 1   & string & & PDB file used for simulation \\
\quad \texttt{psf}     & 1   & string & & Topology file used for simulation \\
\quad \texttt{pdbProteinAtoms} & 1 & string & & PDB file with the $N$ reported atoms \\
\quad \texttt{resid}   & $N$ & integer & & Residue number \\
\quad \texttt{resname} & $N$ & string  & & Residue name \\
\quad \texttt{z}       & $N$ & integer & & Atomic number \\ 
\quad \texttt{.numResidues} & 1 & integer  & &  Number of residues (attribute) \\
\midrule
\quad \texttt{320/} & & & & Group for the 320 K simulations\\
\quad \quad \texttt{0/} & & & & Data of the first replica\\
\quad \quad \quad \texttt{coords} & $F \times N \times 3$ & float  & Å & Atom coordinates \\
\quad \quad \quad \texttt{forces} & $F \times N \times 3$ & float  & kcal/mol/Å & Forces \\
\quad \quad \quad \texttt{dssp}   & $F \times R$          & string &   & DSSP secondary str. assignments \\
\quad \quad \quad \texttt{gyrationRadius} & $F$ & double  & nm & Gyration radius \\
\quad \quad \quad \texttt{rmsd} & $F$ & float  & nm & Root-mean square deviation w.r.t. begin\\
\quad \quad \quad \texttt{rmsf} & $R$ & float  & nm & Cα root-mean-square fluctuation\\
\quad \quad \quad \texttt{box} & $3 \times 3$ & float  & nm & Simulation unit cell \\
\quad \quad \quad \texttt{.numFrames} & 1 & integer  & &  Number of frames for this replica (attribute) \\
\quad \quad \texttt{1/} & & & & Second replica\\
\quad \quad \quad \dots & & & & \\
\midrule
\quad \texttt{348/} & & & & \\
\quad \quad \dots & & & & \\
\bottomrule
\end{tabular}
\caption{Hierarchical organization of the data fields in the mdCATH dataset, with units and description. The following groups and fields are provided in an HDF5 file for each simulated CATH domain. Key: $N$, number of atoms; $R$, number of residues; $F$, trajectory length in frames (1 frame corresponds to 1 ns of simulated time).} 
\label{tab:h5}
\end{table*}

\subsection*{Size}

\begin{figure*}[htbp]
\includegraphics[width=\linewidth]{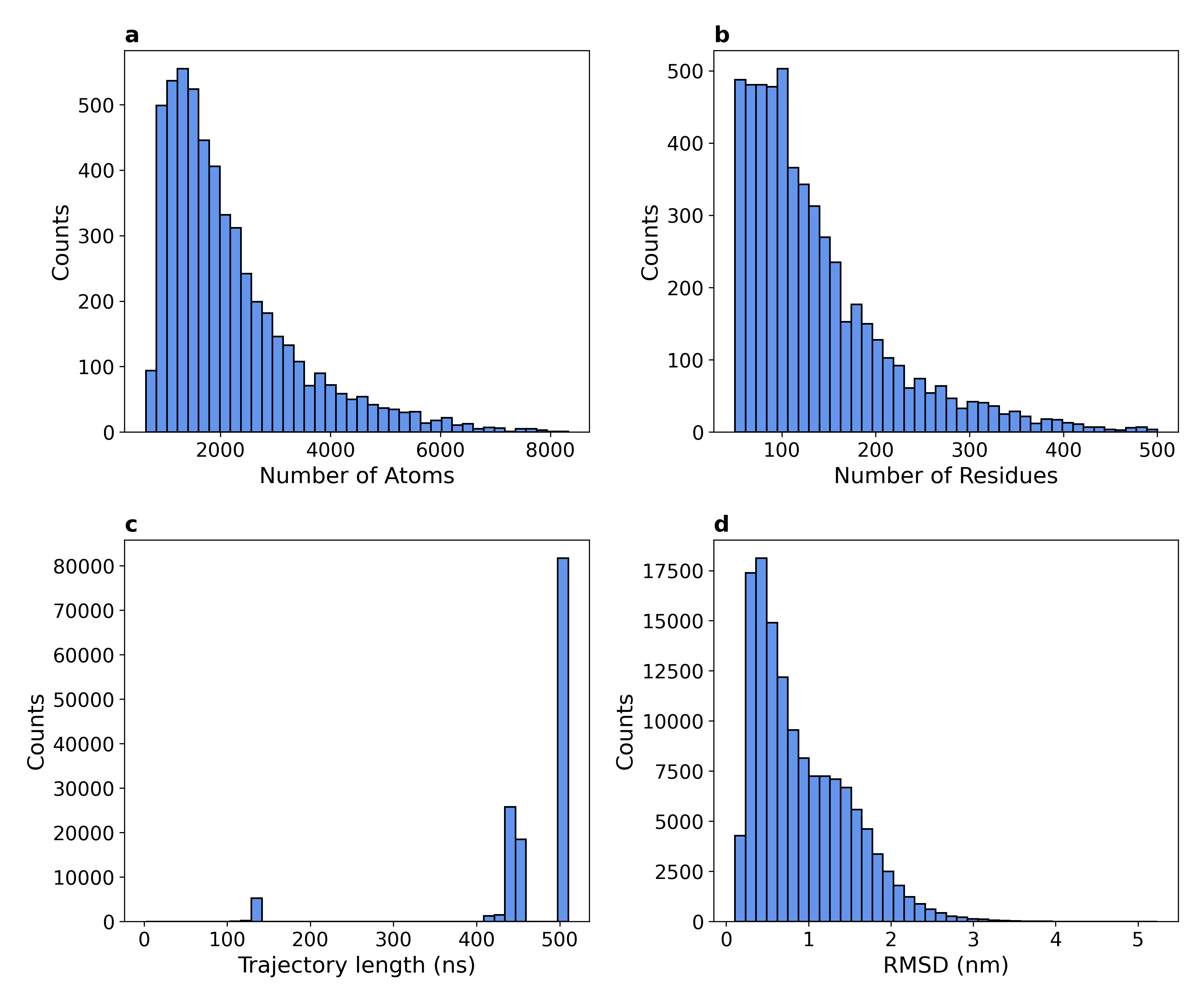}
\caption{\textbf{a}. Distribution of the number of atoms per domain, revealing a variation of nearly an order of magnitude in the atom counts across systems. \textbf{b}. Distribution of the total number of residues per domain, showing a broad peak of around 100 residues and a long tail of up to 500 residues (cut-off size).  \textbf{c}. Distribution of trajectory lengths, peaking at 500 ns. \textbf{d}. Distribution of root mean square deviation (RMSD) of the protein's heavy atoms between the first and the last frame of each trajectory.}
\label{fig:dataset_info}
\end{figure*}

At the production cut-off date, we collected 134,950 trajectories for 5,398 domains, which were included in the dataset. Figures \ref{fig:dataset_info}a and \ref{fig:dataset_info}b show the distribution of system sizes that made it to the production simulation phase in terms of the number of solute atoms and the number of amino acids. Due to the distributed nature of the computing network, the length of the simulations varies (independently from system size), the majority of trajectories being 500 ns long (average 464 ns, standard deviation 76 ns; Figure \ref{fig:dataset_info}c). The total simulated time is over 62 ms. The full dataset size is over 3 TB. Further aggregate statistics are reported in Table \ref{tab: descriptive}.

\begin{table}
    \centering
    \begin{tabular}{lc}
    \toprule
      Domains   & 5,398 \\
      Trajectories   & 134,950 \\
      \midrule
      Total sampled time   & 62.6 ms \\
      Total atoms   & 11,671,592 \\
      Total amino acids   & 740,813 \\
      \midrule
      Avg. traj. length & 464 ns \\
      Avg. system size & 2,162 atoms \\ 
      Avg. domain length & 137 AAs \\ 
      \midrule
      Total file size & 3.3 TB \\
      \bottomrule
    \end{tabular}
    \caption{Descriptive statistics of the mdCATH dataset.}
    \label{tab: descriptive}
\end{table}

\section*{Technical Validation}

We perform several statistical analyses of the dataset to validate its content.

\subsection*{Validation of temperature denaturation}

\begin{figure*}[tbp]
    \centering
    \includegraphics[width=\linewidth]{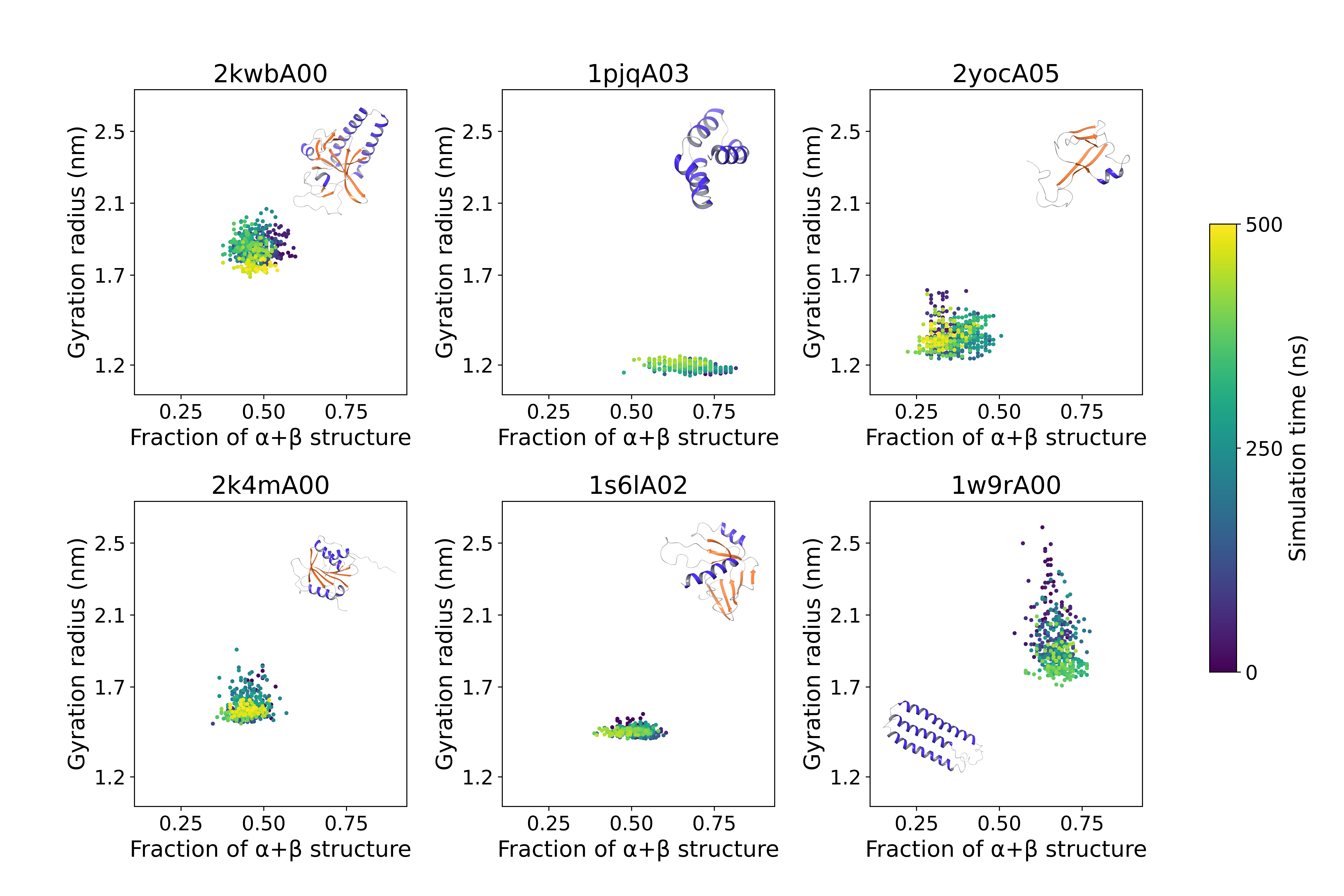}
    \caption{Relation between the radius of gyration and the number of residues in α or β secondary structure elements for six of the mdCATH domains simulated. Each point represents a frame, taken between 0 and 500 ns (blue to yellow) at 1 ns intervals, from the first replica of a run at 320 K.}
    \label{fig:rg-ss-varying-domains}
\end{figure*}

As a first validation of the dataset, we examined the correlation between the amount of secondary structure and the radius of gyration, which was assumed to be a proxy for domain compactness. The fraction of amino acids that are in helical or β-strand configurations, represented by the DSSP codes G, H, I, E, and B, is used to define the amount of secondary structure. This will be referred to as ``α+β'' for simplicity. Figure \ref{fig:rg-ss-varying-domains} shows the results for six domains at 320 K (only one replica is shown for clarity). The radius of gyration and the fraction of sequence in secondary structure elements naturally depend on the domain architecture. At 320 K the domains are generally stable, and both values exhibit fluctuations around mean values but no systematic drift nor marked correlations, with the possible exception of \texttt{1w9rA00}, which undergoes a transition compacting its radius of gyration from 2.4 nm to 1.8 nm.  

\begin{figure*}[tbp]
    \centering
    \includegraphics[width=\linewidth]{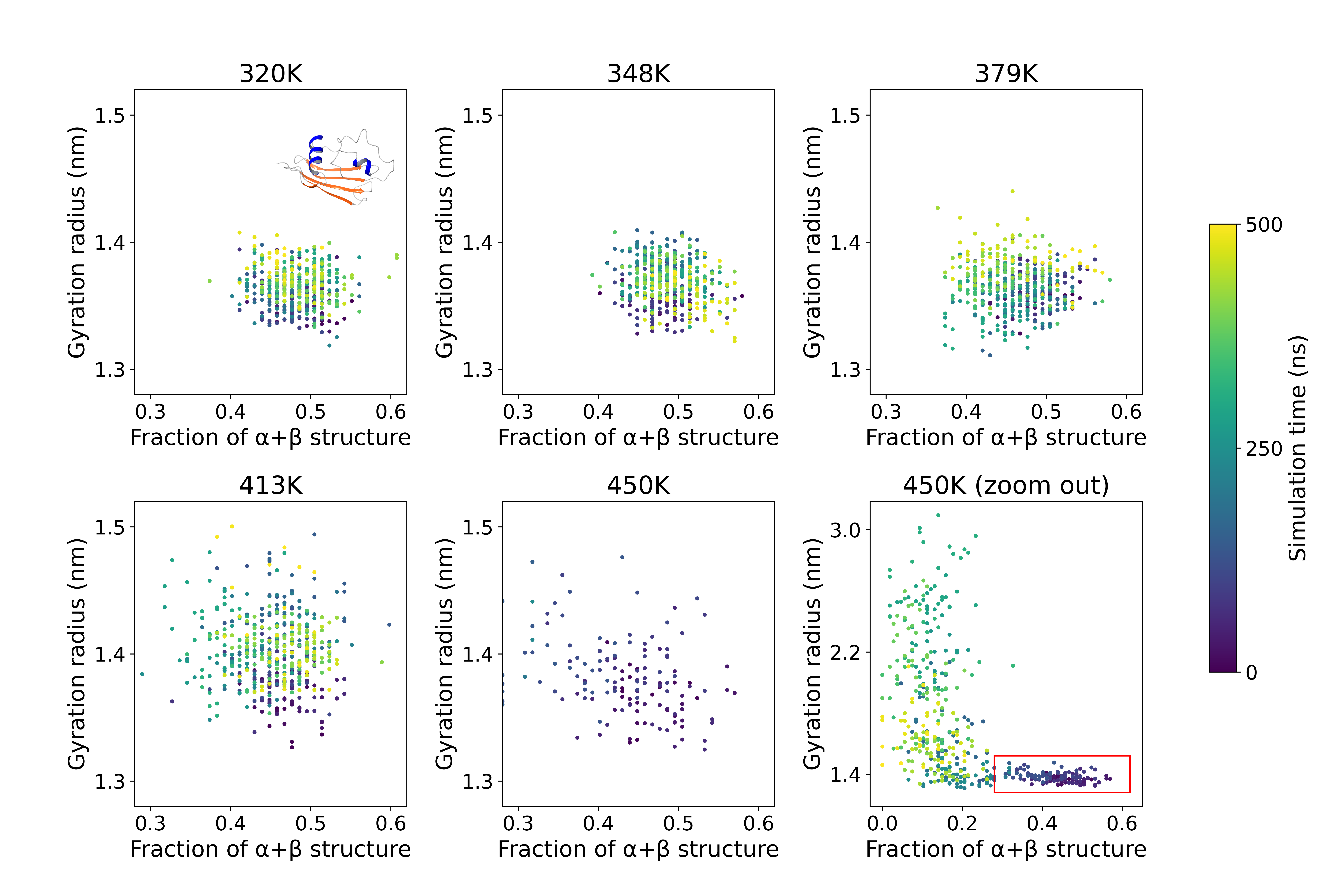}
    \caption{Relation between the radius of gyration and the number of residues in α or β secondary structure elements for domain \texttt{5sicI00} (subtilisin inhibitor-like, a 2-layer α-β sandwich of 106 amino acids) at increasing temperatures. Destabilization is seen at 413~K, and at 450 K complete unfolding occurs within 100 ns (last two panels, fixed scale and full view respectively). Axes and legend are as in Figure \ref{fig:rg-ss-varying-domains}.}
    \label{fig:rg-ss-varying-temperatures}
\end{figure*}

We then validated whether the relationship holds at increasing temperatures. Figure \ref{fig:rg-ss-varying-temperatures} shows the relation between the radius of gyration and the fraction of sequence in secondary structure elements for a specific domain, subtilisin inhibitor-like, a 2-layer α-β sandwich of 106 amino acids (CATH-Gene3D entry G3DSA:3.30.350.10), at increasing temperatures. Between 320 K and 379 K, the dynamics appear essentially unchanged, namely both quantities fluctuate randomly and uncorrelated within the 500 ns of sampled time. Some destabilization starts to appear at 413 K:  the fraction of α/β structure is unchanged, while the radius of gyration has a marked increase beyond the 1.4 nm threshold. At 450 K the system unfolds: the amount of secondary structure drops below 30\%, and the radius of gyration grows beyond 1.5 nm within 100 ns.

\subsection*{Fluctuation-unfolding cooperativity}

\begin{figure}[tbp]
    \centering
    \includegraphics[width=0.7\textwidth]{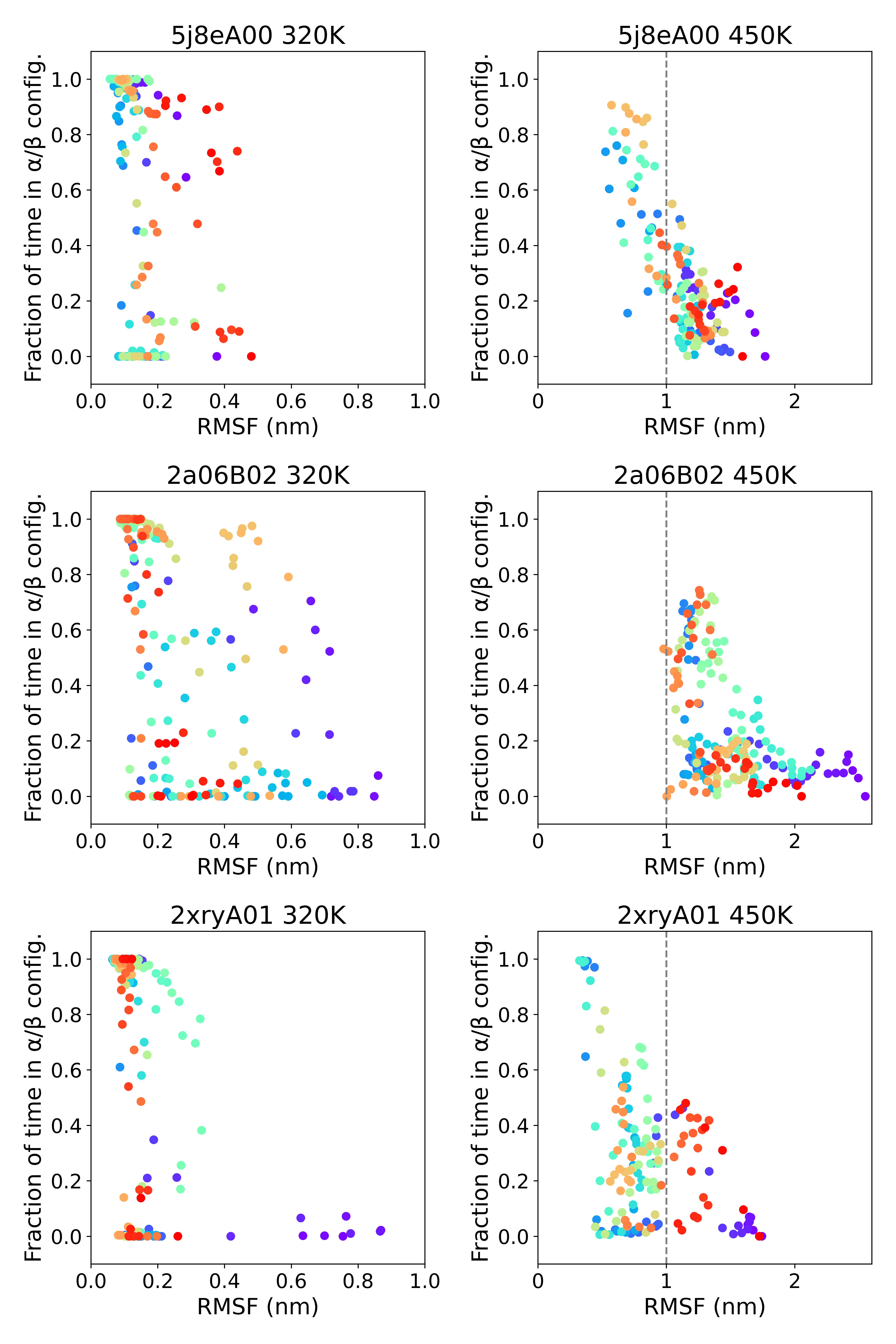}
    \caption{Relationship between the fraction of time each residue spends in α or β secondary structure elements and its root mean squared fluctuation (RMSF) across actin-binding protein T-fimbrin domain (\texttt{5j8eA00}), cytochrome Bc1 complex domain (\texttt{2a06B02}), and HUP superfamily domain (\texttt{2xryA01}). Each point indicates a residue, colored by its position along the sequence, from purple (N-terminal) to red (C-terminal). The relationship is presented in two temperature conditions, 320 K (left) and 450 K (right; note the different RMSF scale). At 320 K, secondary structure presence exhibits a bimodal distribution, weakly correlated with RMSF. Bimodality disappears at 450 K, showing a continuum in the participation to structure elements, which is roughly inversely correlated to the corresponding fluctuations.}
    \label{fig:ss-rmsf}
\end{figure}

We further validated the dataset by assessing the fluctuation of residues in relation to secondary structure and temperatures.  Figure \ref{fig:ss-rmsf} displays, for each residue, the fraction of time spent in an α or β secondary structure element compared to the root mean squared fluctuation (RMSF) of the same residue. The structure-fluctuation relationships are shown for three domains taken as examples, namely \texttt{5j8eA00} (actin-binding protein, T-fimbrin, domain 1; mainly α), \texttt{2a06B02} (cytochrome Bc1 complex, chain A, domain 1; α-β), and \texttt{2xryA01} (HUP superfamily, 6-strand sheet Rossmann fold),  in rows, each shown at low  (320 K, left column) and high temperature (450 K, right column).
%
A clear inverse relationship between local structure and fluctuation emerges which supports that the dataset is well constructed.

\subsection*{Class-wise thermodynamics of denaturation}

It is possible to combine the annotations and metadata provided by the CATH database to cross-reference dynamic data with protein classification. For example, we can leverage CATH metadata by conditioning the analysis on the top-most classification level of CATH (Class), defined in terms of the general architectural organization of the domain: \textit{mainly α}, \textit{mainly β}, \textit{α-β}, \textit{few secondary structures}, and \textit{special}.   

\begin{figure}[tbp]
    \includegraphics[width=0.95\textwidth]{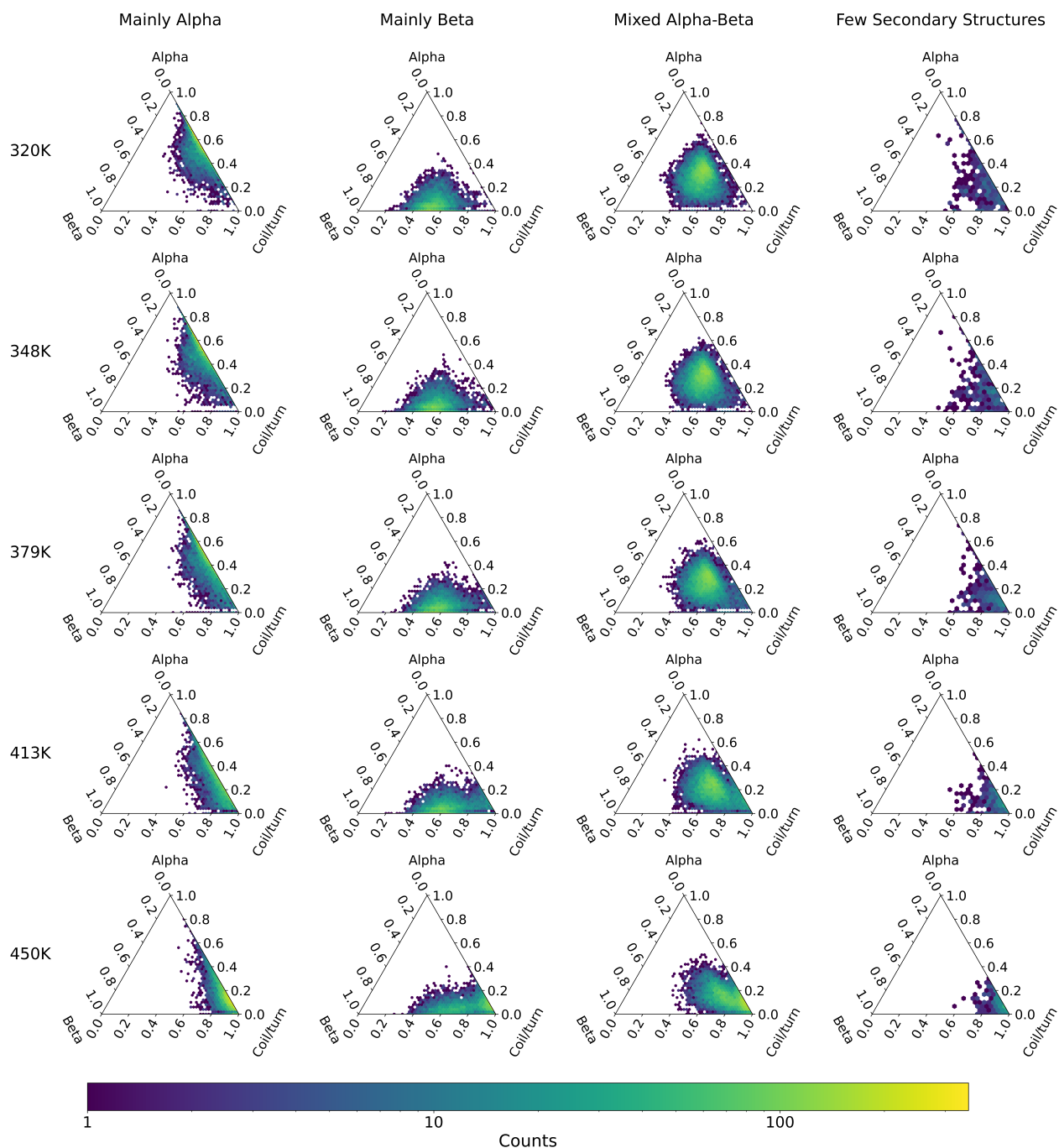}
    \caption{Distribution of protein secondary structures content --- helical (at the top), strand (on the left), and coil/turn (on the right) --- organized by CATH domain class and temperature. It represents data taken from the final snapshot of all replicas across all domains, illustrating how the proportions of helical and strand structures shift toward coil content as temperatures increase.}
    \label{fig:ternary-class-temperature}
\end{figure}

 Figure \ref{fig:ternary-class-temperature} illustrates the construction of probability distribution for various domains, conditioned using domain class annotations. This figure uses ternary plots to show the distribution of protein secondary structures—helical (top), strand (left), and coil/turn (right) content—on a plane. These plots are based on data from the last snapshot of all replicas across all domains, categorized by temperature and domain type. The plots clearly show a shift in the fractions of helical and strand structures toward coil content at temperatures of 413 K and 450 K. Notably, the strand content shows greater resistance to thermal denaturation compared to the helical content.

\subsection*{Kinetics of secondary structure loss}


As a last example, we show how it is possible to combine the annotations and metadata provided by the CATH database to extract proteome-wide kinetic data. Supplementary Figure S1 analyzes the conservation of α/β structure in time as a function of temperature for the four classes (mdCATH has no representative of the ``special'' class). Each panel reports time on the horizontal axis and the fraction of residues in secondary structure elements, normalized so the initial value is one, on the vertical axis. Values for 50 domains per class and replicas are aggregated and displayed as distributions. Different cooperativity regimes emerge for the four classes (Kolmogorov-Smirnov tests for all distribution pairs at 400 ns: $p \ll 10^{-6}$). \textit{Mainly β} domains appear to be the most stable, losing structure only at 450 K. \textit{Mainly α} domains exhibit a partial loss of structure at 413 K;  interestingly, at 450 K their transition to a low-secondary structure state is, on average, abrupt ($\sim 100$ ns). \textit{Mixed α-β} domains have an intermediate behaviour showing aspects of both. Lastly, as expected, the \textit{few secondary structures} class is pretty much diffuse and heterogeneous.

\section*{Usage notes}

An ad-hoc class, \verb+torch_geometric.data.Dataset+, has been integrated into TorchMD-Net\cite{pelaez2024torchmd} to streamline the use of the mdCATH dataset, providing precise control over the protein domain selection and advanced filtering options for trajectories. Listing \ref{lst:torchmd} shows a self-contained code demonstrating how to use the mdCATH data loader in TorchMD-Net for model training, highlighting how additional dataset arguments can be used to focus on specific cases of interest. Future dataset releases will include additional simulations at 300~K to expand coverage around room-temperature conditions.

\begin{lstfloat}
    \centering
    \begin{lstlisting}[language=Python]
import os
import lightning.pytorch as pl
from torchmdnet.data import DataModule
from torchmdnet.module import LNNP
from torchmdnet.scripts.train import get_args

args = get_args()  # default arguments by tmdnet
args = vars(args)  # convert to dictionary
pargs = {
    'dataset': 'MDCATH',
    'dataset_arg': {
      'numAtoms': 3000,
      'numResidues': 250,
      'pdb_list': ['12asA00', '153lA00', '1pytA00'],
      'temperatures': ['320', '348', '379'],
      'skip_frames': 2,
      'solid_ss': 50,
    },
    'log_dir': './logs',
    'model': 'tensornet',
    'num_epochs': 100,
    'embedding_dimension': 64,
    'num_layers': 0,
    'num_rbf': 16,
    'rbf_type': 'expnorm',
    'activation': 'silu',
    'cutoff_lower': 0.0,
    'cutoff_upper': 5.0,
    'max_z': 100,
    'max_num_neighbors': 64,
    'derivative': True
}
# Update the default arguments with the new ones
args.update(pargs)
os.makedirs(args['log_dir'], exist_ok=True)

data = DataModule(args)
data.prepare_data()
data.setup("fit")
lnnp = LNNP(args, 
    prior_model=None, 
    mean=data.mean, 
    std=data.std)
trainer = pl.Trainer(max_epochs=args['num_epochs'])
trainer.fit(lnnp, data)
model = LNNP.load_from_checkpoint(trainer.checkpoint_callback.best_model_path)
trainer = pl.Trainer(inference_mode=False)
trainer.test(model, data)
\end{lstlisting}
    \caption{Importing mdCATH  as a training set in TorchMD-NET.}
    \label{lst:torchmd}
\end{lstfloat}

\begin{lstfloat}
    \centering
    \begin{lstlisting}[language=Python]
from huggingface_hub import HfApi
from huggingface_hub import hf_hub_download

# Initialize the API
api = HfApi()

# Retrieve from HF mdCATH 'data' subdirectory and store in local_dir
hf_hub_download(repo_id="compsciencelab/mdCATH", 
                filename=f"mdcath_dataset_1r9lA02.h5",
                subfolder='data',
                local_dir='.',
                repo_type="dataset")
                    
\end{lstlisting}
    \caption{Example of how to download an mdCATH HDF5 file using the HuggingFace API.}
    \label{lst:hf_api_mdcath}
\end{lstfloat}

\begin{figure}
    \centering
    \includegraphics[width=0.85\textwidth]{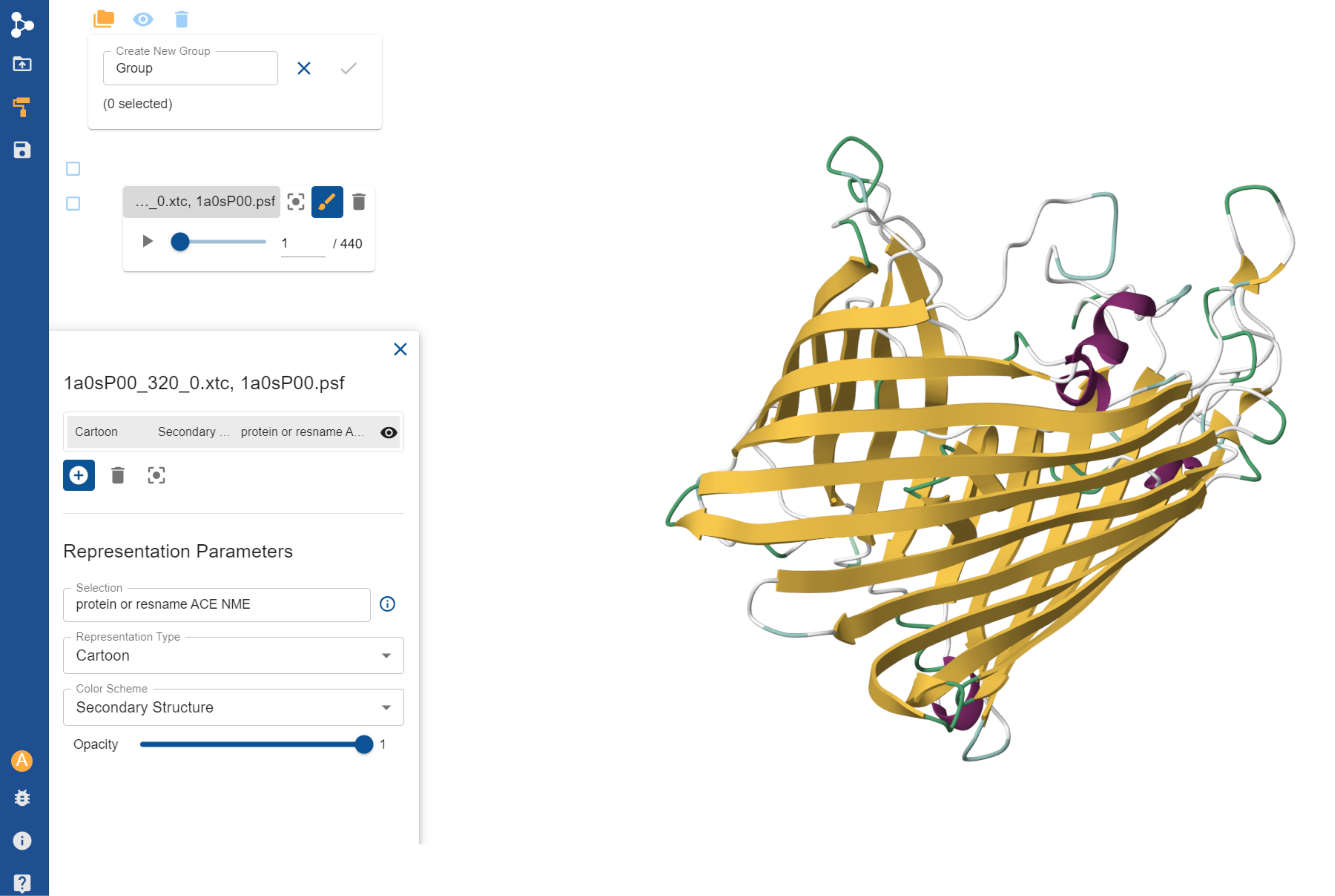}
    \caption{Example of an mdCATH trajectory loaded in the PlayMolecule platform.}
    \label{fig:playmolecule}
\end{figure}

\section*{Code Availability}

Companion code  to load the HDF5 files in VMD \cite{humphrey_vmd_1996} for interactive inspection and analysis, to import them in HTMD molecular analysis library \cite{doerr_htmd_2016}, and to convert them to standard molecular file formats (PDB and XTC) is provided at \url{https://github.com/compsciencelab/mdCATH}. In addition to HuggingFace, the full dataset is also available in the PlayMolecule.org interactive viewer at \url{https://open.playmolecule.org/mdcath}, both for visualization and for further processing via the PlayMolecule platform \cite{torrens-fontanals_playmolecule_2024, martinez-rosell_playmolecule_2017} (Figure \ref{fig:playmolecule}). All the scripts used to generate and analyze the mdCATH dataset are also available at \url{https://github.com/compsciencelab/mdCATH}. 

\vspace{1cm}

\section*{Acknowledgements}
AM is financially supported by Generalitat de Catalunya's Agency for Management of University and Research Grants (AGAUR) PhD grant FI-1-00278 and  PID2020-116564GB-I00 has been funded by MCIN / AEI / 10.13039/501100011033. 
TG acknowledges financial support from the Spoke 7 of the National Centre for HPC, Big Data and Quantum Computing (Centro Nazionale 01 – CN0000013), funded by the European Union -- NextGenerationEU, Mission 4, Component 2, Investment line 1.4, CUP B93C22000620006; from the PRIN 2022 (BioCat4BioPol) from the Ministero dell'Università e Ricerca, funded by the European Union -- NextGenerationEU, Mission 4 Component C2, CUP B53D23015140006; and from the project InvAt-Invecchiamento Attivo e in Salute (FOE 2022) CUP B53C22010140001. 
We thank the volunteers of GPUGRID.net for donating computing time for the simulations.  
Research reported in this publication was partially supported by the National Institute of General Medical Sciences (NIGMS) of the National Institutes of Health under award number R01GM140090. 
The content is solely the responsibility of the authors and does not necessarily represent the official views of the National Institutes of Health.

\section*{Author contributions}
GDF: design and project Lead. TG: generation of the MD data. AM: conversion of MD trajectories into HDF5 datasets. AM, TG and GDF: data analysis and writing-up of the manuscript. 

\section*{Competing Interests}
The authors declare no competing interests.
\bibliography{references}

\end{document}